\documentclass[a4paper]{article}

\usepackage{INTERSPEECH2020}

\title{Continuous Silent Speech Recognition using EEG}
\name{Gautam Krishna, Co Tran, Mason Carnahan, Ahmed H Tewfik}
\address{
  Brain Machine Interface Lab, The University of Texas at Austin}
\email{}

\begin{document}

\maketitle
\begin{abstract}
In this paper we explore continuous silent speech recognition using electroencephalography (EEG) signals. We implemented a connectionist temporal classification (CTC) automatic speech recognition (ASR) model to translate EEG signals recorded in parallel while subjects were reading English sentences in their mind without producing any voice to text. 

Our results demonstrate the feasibility of using EEG signals for performing continuous silent speech recognition. We demonstrate our results for a limited English vocabulary consisting of 30 unique sentences.  
\end{abstract}
\noindent\textbf{Index Terms}: electroencephalography (EEG), silent speech recognition, deep learning, CTC, technology accessibility 

\section{Introduction}

A continuous silent speech recognition model tries to decode what a person was reading in their mind. It can be considered close to mind reading problem where thoughts are also decoded. 
Research along this direction can enable people with severe cognitive disabilities to use virtual assistants like Siri, Alexa, Bixby etc there by improving technology accessibility. It can also enable people with cognitive disabilities to communicate with other people. 
Continuous silent speech recognition technology can also potentially allow soldiers and scientists to perform covert communication in sensitive working environments. Finally continuous silent speech recognition  technology can introduce a new form of thought based communication for able bodied people. 

Electroencephalography (EEG) is a non invasive way of measuring electrical activity of human brain my placing EEG sensors on the scalp of the subject. EEG signals have high temporal resolution even though the spatial resolution is poor. On the other hand Electrocorticography (ECoG) is an invasive way of measuring electrical activity of human brain. ECoG signals have similar temporal resolution like EEG signals but has better spatial resolution and signal to noise ratio (SNR) than EEG signals. The major draw back of ECoG is that it is an invasive procedure requiring the subject to undergo a brain surgery in-order to implant the ECoG electrodes. 
In this work we use non invasive EEG signals to decode the thoughts of the subjects or perform continuous silent speech recognition. 

In \cite{krishna2019speech,krishna20,krishna2019state} authors demonstrated isolated and continuous speech recognition using EEG signals recorded in parallel while subjects were speaking out loud the English sentences and while they were listening to the English utterances for a limited English vocabulary. Authors in \cite{krishna20,krishna2019state,krishna2019speech} used end-to-end automatic speech recognition (ASR) models like connectionist temporal classification (CTC) \cite{graves2014towards}, attention model \cite{chorowski2015attention} and transducer model \cite{graves2013speech} to translate EEG input features directly to text. In a very recent work described in \cite{krishna2020synthesis,krishna2020advancing} authors demonstrated the feasibility of synthesizing speech directly from EEG features. 
Even though in \cite{krishna2019state} authors demonstrated speech recognition using EEG signals recorded during passive listening their experiments didn't involve subjects explicitly reading sentences in their mind. Hence it is not clear whether the work described in \cite{krishna2019state} studies continuous silent speech recognition problem. 
In this work we perform continuous silent speech recognition where we use a CTC model to map EEG features recorded while the subjects were reading English sentences in their mind, to text.  

Other related works include \cite{kumar2018envisioned} where authors demonstrated envisioned speech recognition using random forest classifier and in \cite{wang2017simulation} authors demonstrated imagined speech recognition from EEG signals using synthetic EEG data and CTC network but in our work we use real experimental EEG data and larger vocabulary. In \cite{ramsey2017decoding} authors demonstrated speech recognition using ECoG signals. In \cite{zhao2015classifying} the authors used classification approach for identifying phonological categories in imagined and silent speech but in this paper we demonstrate continuous silent speech recognition. In \cite{porbadnigk2009eeg} authors demonstrated EEG based silent speech recognition for a vocabulary of five words but not at sentence level where continuous recognition is performed. Also in \cite{porbadnigk2009eeg} authors used traditional hidden markov model (HMM) model but in this work we make use of state-of-the-art deep learning models and results are demonstrated for a much larger vocabulary size. In \cite{kapur2018alterego,wadkins2019continuous} authors perform silent speech recognition but they didn't use EEG neural recordings, in our work we make of EEG neural recordings which can lead to future work on mind reading or decoding thoughts. 
Similarly in \cite{kapur2018alterego} authors didn't demonstrate continuous speech recognition and the work described in \cite{wadkins2019continuous} is closely related to our work but they didn't use EEG features for decoding and moreover our approach differs from them as they make use of convolutional neural network (CNN) to extract features whereas we extract more interpretable hand craft EEG features\cite{krishna2019speech,krishna20} to train the model. 

The major contribution of this work is the demonstration of feasibility of using EEG features to perform continuous silent speech recognition. We believe our results will motivate the research community to improve our results and come up with better state-of-the-art models that can perform continuous silent speech recognition using EEG features. 

\section{Connectionist Temporal Classification (CTC)}

The CTC ASR model ideas were first introduced in \cite{graves2006connectionist,graves2014towards}. The CTC model can perform continuous speech recognition by making the length of output tokens equal to number of time steps of the input features by allowing repetition of output tokens and by introducing a special token called blank token. Thus the CTC model is alignment free.  The CTC ASR model consists of an encoder, decoder and a CTC loss function.  

The encoder of our CTC model consists of two layers of gated recurrent unit (GRU) \cite{chung2014empirical} with 128 hidden units in first GRU layer and 64 hidden units in the second GRU layer. Each GRU layer had a dropout regularization \cite{srivastava2014dropout} with a dropout rate 0.1. The GRU layers were followed by a temporal convolutional network (TCN) \cite{bai2018empirical} consisting of 32 filters. The decoder of the CTC model consists of a time distributed dense layer and a softmax activation function. The output of the encoder is fed into the decoder at every time step. The encoder takes the EEG features as input. The number of time steps of the encoder is calculated as the product of sampling frequency of the input EEG features and sequence length. There was no fixed value for the number of time steps. We used dynamic recurrent neural network (RNN) cell. 

The CTC model was trained for 130 epochs with a batch size of 32 using adam \cite{kingma2014adam} optimizer to optimize the CTC loss function. The mathematical details of CTC loss function are covered in \cite{graves2014towards,graves2006connectionist,krishna2019state,krishna20}. We used a character based model in this work. The model was predicting a character at every time step. During inference time a CTC beam search decoder is used in combination with an external 4-gram language model \cite{toshniwal2018comparison} known popularly as shallow fusion. 
 
We used 80 \% of the total EEG data as training set and remaining data as test set. The train-test split was done randomly. There was no overlap between the training and testing set.  The validation split hyper parameter was set to a value of 0.1. 

Figure 1 shows the architecture of our CTC model and Figure 2 shows CTC training loss convergence. All the scripts were written using Tensorflow 2.0 and Keras deep learning framework.

\begin{figure}[h]
\label{fig:asrmodel}
\includegraphics[height=8.5cm, width=\linewidth,trim={0.1cm 0.1cm 0.1cm 0.1cm}]{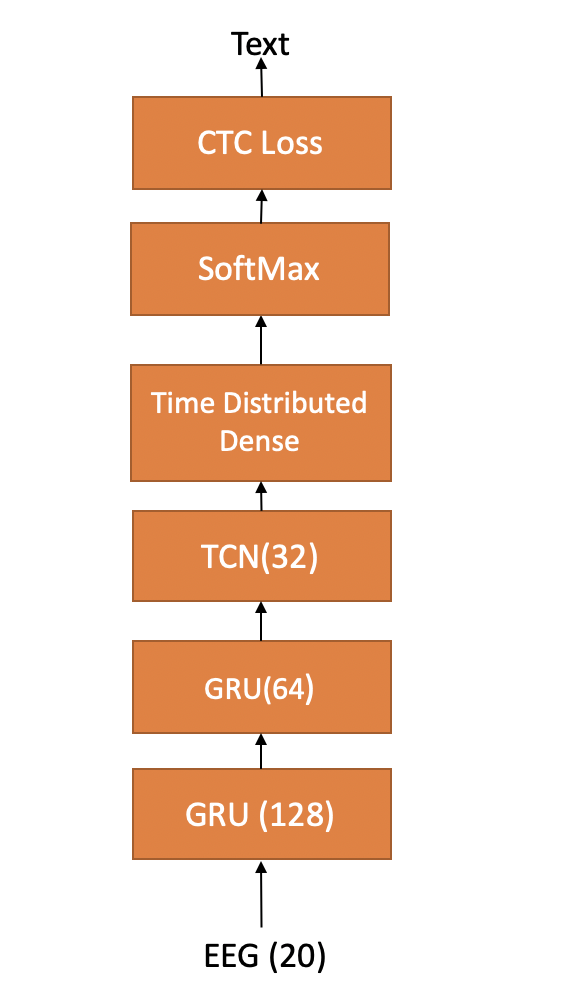}
\caption{CTC ASR Model} 
\label{1vsall}
\end{figure}

\begin{figure}[h]
\begin{center}
\includegraphics[height=5cm, width=0.4
\textwidth,trim={0.1cm 0.1cm 0.1cm 0.1cm},clip]{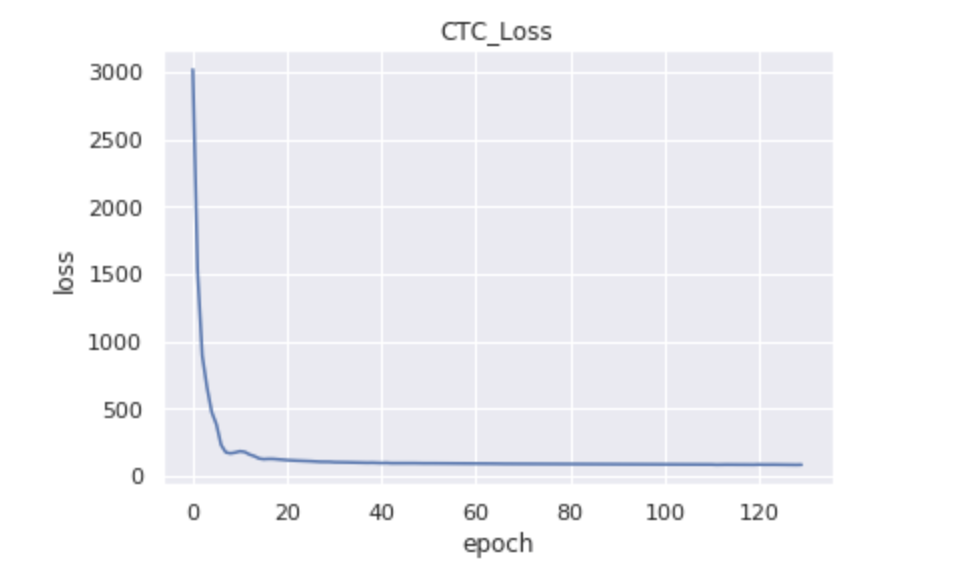}
\caption{CTC loss convergence} 
\label{1vsall}
\end{center}
\end{figure}

\begin{figure}[h]
\begin{center}
\includegraphics[height=3cm,width=0.25\textwidth,trim={1cm 1cm 1cm 0.1cm},clip]{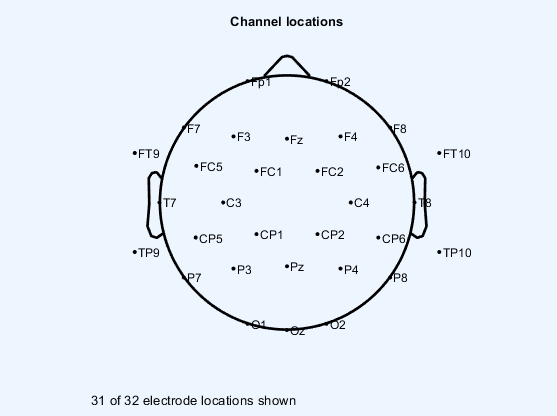}
\caption{EEG channel locations for the cap used in our experiments} 
\label{1vsall}
\end{center}
\end{figure}

\section{Design of Experiments for building the database}

Four male subjects in their early to mid twenties took part in the EEG experiment. Out of the four subjects three were non native English speakers and one subject was a native English speaker. Each subject was asked to read first 30 English sentences from USC-TIMIT database \cite{narayanan2014real} in their mind without producing any voice and their EEG signals were recorded. The English sentences were shown to them on a computer screen. Each subject was then asked to repeat the same experiment two more times. The data was recorded in absence of background noise. There were 90 EEG recordings per each subject.

We used Brain product's EEG recording hardware. The EEG cap had 32 wet EEG electrodes including one electrode as ground as shown in Figure 3. We used EEGLab \cite{delorme2004eeglab} to obtain the EEG sensor location mapping. It is based on standard 10-20 EEG sensor placement method for 32 electrodes. 

\section{EEG feature extraction details}

We followed the same preprocessing methods used by authors in \cite{krishna20,krishna2019speech} to process the EEG data and extract EEG features. 

EEG signals were sampled at 1000Hz and a fourth order IIR band pass filter with cut off frequencies 0.1Hz and 70Hz was applied. A notch filter with cut off frequency 60 Hz was used to remove the power line noise.
The EEGlab's \cite{delorme2004eeglab} Independent component analysis (ICA) toolbox was used to remove other biological signal artifacts like electrocardiography (ECG), electrooculography (EOG) and electromyography (EMG) due to subvocalization from the EEG signals. 
We extracted five statistical features for EEG, namely root mean square, zero crossing rate,moving window average,kurtosis and power spectral entropy \cite{krishna2019speech,krishna20}. Thus in total we extracted 31(channels) X 5 or 155 features for EEG signals. The EEG features were extracted at a sampling frequency of 100Hz for each EEG channel.

\begin{figure}[h]
\centering
\includegraphics[height=5cm, width=0.4
\textwidth,trim={0.1cm 0.1cm 0.1cm 0.1cm},clip]{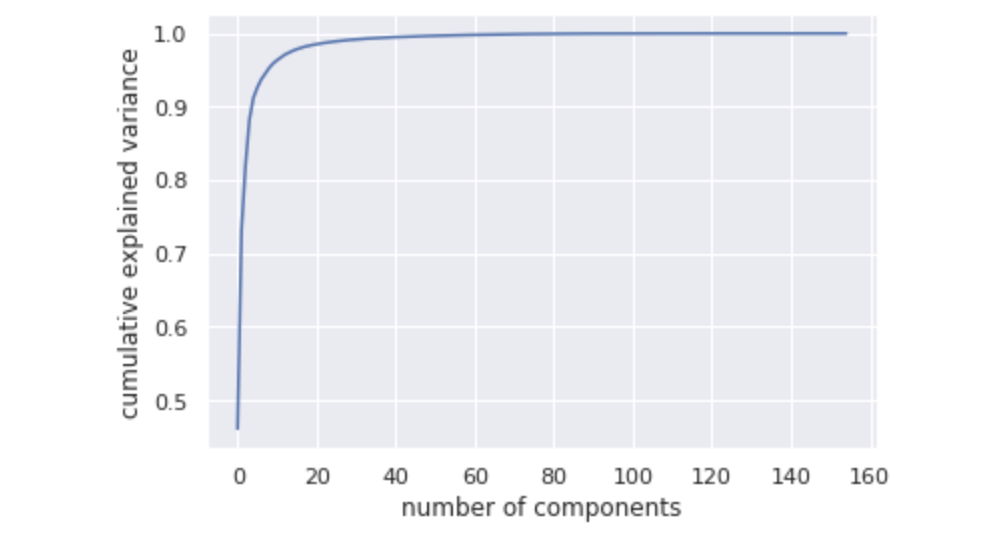}
\caption{Explained variance plot}
\label{1vsall}
\end{figure}

\section{EEG Feature Dimension Reduction Algorithm Details}

After extracting EEG features  as explained in the previous section, we used non linear methods to denoise the EEG feature space \cite{krishna20,krishna2019state}.

We reduced the 155 EEG features to a dimension of 20 by applying Kernel Principle Component Analysis (KPCA) \cite{mika1999kernel} where as in \cite{krishna20,krishna2019state} authors reduced the EEG features to a final dimension of 30 for performing speech recognition using EEG features recorded in parallel with speech or passive listening. We plotted cumulative explained variance versus number of components to identify the right feature dimension as shown in Figure 4. We used KPCA with polynomial kernel of degree 3 \cite{krishna2019speech,krishna20}.

\section{Results}
\label{sec:print}
We used word error rate (WER) as the performance metric to evaluate the CTC model during test time. Table 1 shows the results obtained during test time for different test set vocabulary sizes. The average WER is reported in Table 1. These results were obtained when we used randomly 20 \% of the total data as test set. 

For all the test set vocabulary sizes we observed WER in 70's or 80's (\%). We believe the test time performance can be improved by training the model with more number of examples. We also observed that the WER went up as test set vocabulary size increase except for vocabulary size consisting of 10 unique sentences. Normally as the vocabulary size increase the model need to be trained with more number of examples to give more accurate predictions with better generalization. We believe for test set vocabulary consisting of 10 unique sentences the model reported higher WER compared to larger vocabulary sizes possibly because some words present in the predicted text were not corrected by the external language model resulting in increased WER.   
Usually the CTC is model is trained with larger data sets to observe state-of-the-art performance during test time \cite{graves2014towards}. 

We further observed that our overall results were poor compared to the results demonstrated by authors in \cite{krishna2019improving} where they used EEG signals recorded in parallel with spoken speech and where they demonstrated results for larger vocabulary size during test time. In \cite{krishna2019improving} authors didn't perform silent speech recognition. 
Our overall results indicate that continuous silent speech recognition remains as a challenging problem.

In \cite{gruber2001effects,girbau2007neurocognitive,burgess1999memory} authors show that from neuroscience perspective, during silent speech there is brain activity observed in left frontal lobe due to motor planning for inner speech and activity in inferior, middle frontal gyrus, inferior parietal gyrus during subvocal rehearsal. The silent reading also activates auditory cortex or temporal lobe part superior temporal gyrus (STG) due to self perception involved during reading process. Hence we also performed continuous silent speech recognition using all EEG features from temporal lobe sensors (T7,T8,TP9,TP10) of dimension 20 ( 5 features per channel or sensor) and observed a WER of 86.73 \% during test time for vocabulary consisting of 72 total sentences or 30 unique sentences. We observed a WER of 85.04 \% during test time for the same test set vocabulary when we performed experiment using all frontal lobe (F3,F4,F7,F8,FC1,FC2,FC5,Fp1,Fp2,FT9,FT10,Fz) EEG features of dimension 65. When we performed experiments by combining EEG features from both temporal and frontal lobe of dimension 85 we observed a test time WER of 84.22 \% for the same test set vocabulary. However all these WER's were higher compared to the result we obtained when we used EEG features from all 31 EEG sensors followed by dimension reduction as seen from Table 1. From Table 1 we can see that by using features from all 31 EEG sensors we were able to achieve a test time WER of 83.34 \% for 72 total sentences or 30 unique sentences. The lower the WER value the better the speech recognition inference performance in general. 

We also did continuous silent speech recognition experiment where we trained the CTC model using EEG data from the first three subjects and used the last subject or fourth subject data as test set and observed a higher test time WER of 92.55 \% for test set consisting of 30 unique sentences. This shows that transferability of the model among subjects seems to be challenging for the task for the task of continuous silent speech recognition using EEG.  

 To the best of our knowledge this is the first time continuous silent speech recognition is demonstrated using real experimental EEG features at sentence level. We hope our results will motivate other researchers to develop better decoding models. 

\begin{table*}[!ht]
\centering
\begin{tabular}{|l|l|l|l|l|l|}
\hline
\textbf{\begin{tabular}[c]{@{}l@{}}Total \\ Number\\ of \\ Sentences\end{tabular}} & \textbf{\begin{tabular}[c]{@{}l@{}}Number\\  of\\ Unique\\ Sentences\\ Contained\end{tabular}} & \textbf{\begin{tabular}[c]{@{}l@{}}Total\\ Number\\ of \\ words\\ Contained\end{tabular}} & \textbf{\begin{tabular}[c]{@{}l@{}}Number\\ of\\ Unique\\ words\\ Contained\end{tabular}} & \textbf{\begin{tabular}[c]{@{}l@{}}Number\\ of \\ Letters\\ Contained\end{tabular}} & \textbf{\begin{tabular}[c]{@{}l@{}}EEG\\ WER\\ (\%)\end{tabular}} \\ \hline
12                                                                                 & 5                                                                                              & 79                                                                                        & 29                                                                                        & 343                                                                                 & 74.86                                                             \\ \hline
24                                                                                 & 10                                                                                             & 148                                                                                       & 59                                                                                        & 683                                                                                 & 84.08                                                             \\ \hline
36                                                                                 & 15                                                                                             & 224                                                                                       & 84                                                                                        & 986                                                                                 & 79.63                                                             \\ \hline
48                                                                                 & 20                                                                                             & 301                                                                                       & 106                                                                                       & 1347                                                                                & 81.06                                                             \\ \hline
60                                                                                 & 25                                                                                             & 376                                                                                       & 132                                                                                       & 1740                                                                                & 82.80                                                            \\ \hline
72                                                                                 & 30                                                                                             & 435                                                                                       & 153                                                                                       & 2054                                                                                & 83.34                                                             \\ \hline
\end{tabular}
\caption{WER on test set}
\end{table*}


\section{Conclusion}
\label{sec:refs}
In this work we demonstrated the feasibility of using EEG features to perform continuous silent speech recognition. To the best of our knowledge this is the first time a continuous silent speech recognition using EEG features is demonstrated for predicting sentences. 

Future work will focus on improving our current results by doing EEG source localization, developing better decoding models and training the model with larger EEG data set.

\section{Acknowledgement}
We would like to thank Kerry Loader and Rezwanul Kabir from Dell, Austin, TX for donating us the GPU to train the models used in this work.

\bibliographystyle{IEEEtran}

\bibliography{mybib}


\end{document}